\documentclass{ws-ijmpe}
\usepackage[super,compress]{cite}
\begin{document}
\markboth{G. Saxena \textit{et al.}}{Structural properties and decay modes of Z $=$ 122, 120 and 118 superheavy nuclei}

\catchline{}{}{}{}{}

\title{\textsc{Structural properties and decay modes of Z $=$ 122, 120 and 118 superheavy nuclei}}

\author{G. Saxena}

\address{Department of Physics, Govt. Women Engineering College, Ajmer-305002, India\\
gauravphy@gmail.com}

\author{M. Kumawat}

\address{Department of Physics, School of Basic Sciences, Manipal University, Jaipur-303007, India}

\author{S. Somorendro Singh}

\address{Department of Physics and Astrophysics, University of Delhi,
Delhi 110 007, India}

\author{Mamta Aggarwal}

\address{Department of Physics, University of Mumbai, Kalina Campus, Mumbai-400098, India\\
mamta.a4@gmail.com}
\maketitle

\begin{history}
\received{Day Month Year}
\revised{Day Month Year}
\end{history}

\begin{abstract}
Structural properties and the decay modes of the superheavy elements Z $=$ 122, 120, 118 are studied in a microscopic framework. We evaluate the binding energy, one- and two- proton and neutron separation energy, shell correction and density profile of even and odd isotopes of Z $=$ 122, 120, 118 (284 $\leq$ A $\leq$ 352) which show a reasonable match with FRDM results and the available experimental data. Equillibrium  shape and deformation of the superheavy region are predicted. We investigate the possible decay modes of this region specifically $\alpha$-decay, spontaneous fission (SF) and the $\beta$-decay  and evaluate the probable $\alpha$-decay chains. The phenomena of bubble like structure in the charge density is predicted in $^{330}$122, $^{292,328}$120 and $^{326}$118 with significant depletion fraction around 20-24$\%$ which increases with increasing Coulomb energy and diminishes with increasing isospin (N$-$Z) values exhibiting the fact that the coloumb forces are the main driving force in the central depletion in superheavy systems. \end{abstract}


\ccode{PACS numbers: 23.50.+z, 21.10.-k, 21.10.Dr}


\section{Introduction}
In continuation to our recent study \cite{saxenaijmpe2018} on the superheavy nuclei in Z $=$ 121 region where some distinct features were predicted in  even and odd isotopes of Z $=$ 121 (281 $\leq$ A $\leq$ 380)\cite{saxenaijmpe2018}, we explore further in the unknown territory of superheavy region of Z $=$ 122, 120 and 118 (284 $\leq$ A $\leq$ 352) where the experimental efforts towards the synthesis and study of decay modes of these nuclei are going on  \cite{manju2018,fanli,santosh2018}. Some efforts  have already been made in nuclei with Z $=$ 118, 119, 120 \cite{Hofmann2016,Devaraja2016,Oganessian2009}. The search for the possible fusion reactions with the help of available theoretical data on the distinct features and predictions has been going on at GSI \cite{Hofmann2000,Hofmann2011}, RIKEN \cite{Morita2007} and JINR \cite{Oganessian2010,Oganessian2015,Hamilton2013} that are expected to provide useful information in this domain of periodic chart. Various theoretical attempts \cite{Bao2015,Wang2015,Niyti2015,Heenen2015,Santhosh2016,Budaca2016,Liu2017,Zhang2017,TLZhao2018,JPCui2018}, in particular, the studies of Z $=$ 118 \cite{Sobiczewskiz2016}, Z $=$ 118$-$121 \cite{ShanZhang2017}, Z $=$ 119, 120 \cite{Poenaru2017}, Z $=$ 100$-$126 \cite{Pomorski2018}, Z $=$ 105 $\leq$ Z $\leq$ 135 \cite{Santhosh2018}, Z $=$ 121 \cite{Santhosh2016,KPSanthosh2016}, Z $=$ 122 \cite{Manjunatha2016}, Z $=$ 120, 124 \cite{Mehta2015}, Z $=$ 123 \cite{Santhosh2016a}, Z $=$ 122, 124, 126, 128 \cite{Santhosh2016xrd}, Z $=$ 124, 126 \cite{Manjunatha2016bbm,Manjunatha2016zia}, Z $=$ 132, 138 \cite{Rather2016} have significantly contributed to the knowledge of this unexplored region of nuclear landscape. In addition, possible new shell closures \cite{Wu2003,Zhang2004,Adamian2009,Biswal2014,Brodzinski2013,Moller1994,Rutz1996,Cwiok1996}, the exotic phenomenon of bubble/semi-bubble structures \cite{Berger2001,Decharge2003,Afanasjev2005,SinghSK2013} and the possible decay modes \cite{Hofmann2016,Oganessian2009,Bao2015,Wang2015,Niyti2015,Heenen2015,Santhosh2016,Budaca2016,Liu2017,Zhang2017,Ikram2017}are being anticipated in this superheavy region which makes the study of this region very interesting that might open up new avenues for nuclear physics research.

On the experimental front, the most probable projectile-target combinations to synthesize these nuclei (Z $=$ 122 isotopes) have been reported as Cr+Cf, Fe+Cm, Se+Ra, and As+Ac \cite{manju2018}. To produce superheavy nuclei with Z $=$ 120 the system $^{40}$Ca + $^{257}$Fm is found most favorable with maximal production cross sections (optimal incident energies) of 1.24 pb (205.66 MeV) \cite{fanli}. The theoretical prediction on the synthesis of superheavy nuclei with Z $=$ 118 ($^{290-302}$Og) using $^{48}$Ca, $^{45}$Sc, $^{50}$Ti, $^{51}$V, $^{54}$Cr, $^{55}$Mn, $^{58}$Fe, $^{59}$Co and $^{64}$Ni induced reactions have been made recently \cite{santosh2018} along with most probable projectile-target combinations \cite{manjunpa2018}. Despite the synthesis of nuclei as heavy as $^{294}$118, there are plenty of canvasses which are yet to be explored to alleviate experimental attempts and to embark in this new arena of superheavy nuclei, this systematic study has been planned.\par
We use our well established theoretical formalisms (i) Relativistic mean-field plus state dependent BCS (RMF+BCS) approach \cite{Serot1984,Ring1996,Yadav2004,Saxena2017,Saxena2017hzo,Saxenaplb2018} and (ii) Macroscopic-Microscopic approach with triaxially deformed Nilsson Strutinsky method (NSM) \cite{Aggarwal2010,Aggarwal2014} which have proved to be simple yet very effective to investigate exotic unknown regions of periodic chart quite well. The presentation of this work is threefold: (i) Prediction of the proton drip line and the ground state properties namely the binding energy, separation energy, shell correction, deformation and shape (ii) evaluation of charge density and neutron density to perform a systematic study of central density depletion, if present, in this region and (iii) the investigation of possible decay modes specifically $\alpha$-decay, spontaneous fission (SF) and $\beta$-decay where we also present the possible $\alpha$-decay chains of the nuclei of interest in this work. Our calculations provide significant inputs about the ground state properties and magicity in N $=$ 164 and 228, existence of bubble structure and possible decay modes. Among Z $=$ 122 superheavy nuclei, alpha decay has been found most dominant decay mode for $^{307-314}$122.  A good agreement with the available experimental data \cite{Wang-Mass2017} and Finite Range Droplet Model (FRDM) calculations \cite{Moller2012} proves the reliability of the theoretical formalisms used in this work.\par

\section{Theoretical Formalisms}
The calculations in the relativistic mean-field plus state dependent BCS (RMF+BCS) approach \cite{Serot1984,Ring1996,Yadav2004,Saxena2017,Saxena2017hzo,Saxenaplb2018} have been carried out using the model Lagrangian density with nonlinear terms both for the ${\sigma}$ and ${\omega}$ mesons along with TMA parametrization \cite{Singh2013,Yadav2004}. The corresponding Dirac equations for nucleons and Klein-Gordon equations for mesons obtained with the mean-field approximation are solved by the expansion method on the widely used axially deformed Harmonic-Oscillator basis \cite{Geng2003,Gambhir1989}. The quadrupole constrained calculations have been performed for all the nuclei considered here in order to obtain their potential energy surfaces (PESs) and determine the corresponding ground-state deformations \cite{Geng2003,Flocard1973}. For nuclei with odd number of nucleons, a simple blocking method without breaking the time-reversal symmetry is adopted \cite{Geng2003wt,Ring1996}. In the calculations we use for the pairing interaction a delta force, i.e., V $=$ -V$_0 \delta(r)$ with the strength V$_0$ $=$ 350 MeV fm$^3$ which has been used in Refs.$~$ \cite{Yadav2004,Saxena2017} for the successful description of drip-line nuclei. Apart from its simplicity, the applicability and justification of using such a $\delta$-function form of interaction has been discussed in Ref.$~$\cite{Dobaczewski1983}, whereby it has been shown in the context of HFB calculations that the use of a delta force in a finite space simulates the effect of finite range interaction in a phenomenological manner \cite{Bertsch1991}. For further details of these formulations we refer the reader to Refs.$~$\cite{saxenaijmpe2018,Gambhir1989,Singh2013,Geng2003}. \par

Macroscopic-Microscopic approach using the triaxially deformed Nilsson Strutinsky method (NSM) treats the structural properties of the atomic nuclei which are governed by the delicate interplay of macroscopic bulk properties of the nuclear matter and the microscopic shell effects. The microscopic effects arising due to nonuniform distribution of nucleons are included through the  Strutinsky's shell correction $\delta$E$_{shell}$ ~\cite{Strutinsky1968,BRACK1972,Nilsson1972} along with the deformation energy E$_{def}$ (obtained from the surface and Coulomb effects). The shell correction to energy  $\delta$E$_{shell}$ can be written as $\delta E_{shell}$=$\sum_{i=1}^A \epsilon_i- \tilde E$, where the first term is the shell model energy in the ground state and the second term is the smoothed energy with the smearing width 1.2$\hbar$ $\omega$. The energy E ($=$ -BE) is minimized with respect to nilsson deformation and shape parameters ($\beta$ and $\gamma$ where the axial deformation parameter $\beta$ varies from 0 to 0.4 in steps of 0.01. The angular deformation parameter $\gamma$, which is used to evaluate shapes, varies from -180$^o$ (oblate shape) to -120$^o$ (prolate shape) and -180$^o$ $<$ $\gamma$ $<$ -120$^o$ (triaxial shapes)). The $\beta$ and $\gamma$ corresponding to E minima give the deformation and shape of the nucleus. For further details on NSM method, we refer the reader to Refs.$~$\cite{saxenaijmpe2018,Aggarwal2010,Aggarwal2014} \par

\section{Results and discussions}
In the first subsection, we present our results of ground state properties of Z $=$ 122, 120 and 118 superheavy nuclei (284 $\leq$ A $\leq$ 352) and show a comparison between the two considered theories RMF and NSM along with results of Finite Range Droplet Model (FRDM). We explore possible new magic and bubble like structures in this region. Thereafter, in the next subsection, we investigate possible decay modes such as $\alpha$-decay, spontaneous fission (SF) and $\beta$-decay in these nuclei for full isotopic chains and subsequently report possible $\alpha$-decay chains and their half lives along with available experimental data at places for comparison.\par

\subsection{Ground State Properties}

\begin{figure}[h]
\centering
\includegraphics[width=0.7\textwidth]{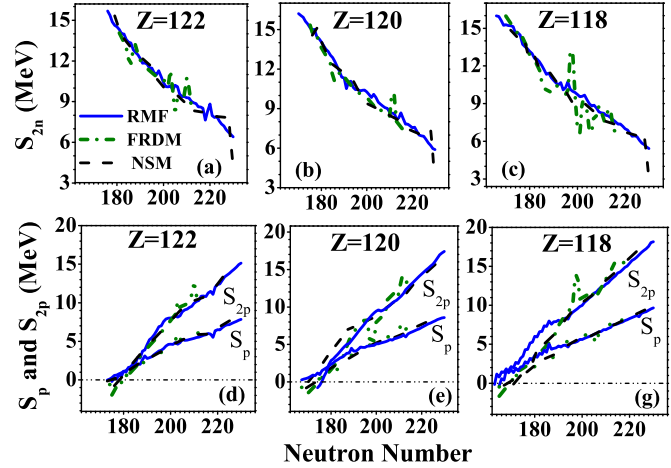}
\caption{(Colour online) Variation of two neutron separation energy (S$_{2n}$), one- and two-proton separation energies (S$_p$ and S$_{2p}$) with neutron number N for Z $=$ 122, 120 and 118 isotopes. The onset of proton drip line is shown.}
\label{fig1}
\end{figure}

Fig. \ref{fig1} shows the variation of two neutron separation energy S$_{2n}$ (upper panels) with neutron number N for Z $=$ 122, 120 and 118 isotopes, which is calculated using RMF approach \cite{Yadav2004,Saxena2017} and Mac-Mic approach with NS prescription (NSM) \cite{Aggarwal2010,Aggarwal2014}. For a comparison, FRDM data values \cite{Moller2012} are also plotted. One can see that all these three approaches (RMF, NSM and FRDM) are in reasonable agreement for this superheavy region and validate our calculations. We have calculated one- and two-proton separation energies (S$_p$ and S$_{2p}$) for these chains to identify the first one- and two-proton unbound nuclei which indicates the onset of 1p and 2p drip lines (shown in lower panels of Fig. \ref{fig1}). The last 2p and 1p bound nuclei are found (using RMF) to be $^{298}$122 \& $^{296}$122 for Z $=$ 122, $^{290}$120 and  $^{287}$120 for Z $=$ 120 and $^{284}$118 \& $^{282}$118 for Z $=$ 118 respectively. However, from NSM, we found these results in a little variance which gives $^{301}$122, $^{296}$120, $^{288}$118 \& $^{296}$122, $^{294}$120, $^{286}$118 as last 2p- and 1p-bound candidates for Z $=$ 122, 120 and 118, respectively. One can also notice that in case of FRDM for Z $=$ 120 and Z $=$ 118 there are jumps in two neutron separation energy S$_{2n}$ and two proton separation energy S$_{2p}$ for N $=$ 198 which is found due to spurious shell closure at N $=$ 198 \cite{ahmad}. The Binding energy per nucleon curve for all these isotopic chains are found in the perfect shape of parabola (not shown here) and after examining those we restrict our calculation only upto N $=$ 230.\par

\begin{figure}[h]
\centering
\includegraphics[width=0.7\textwidth]{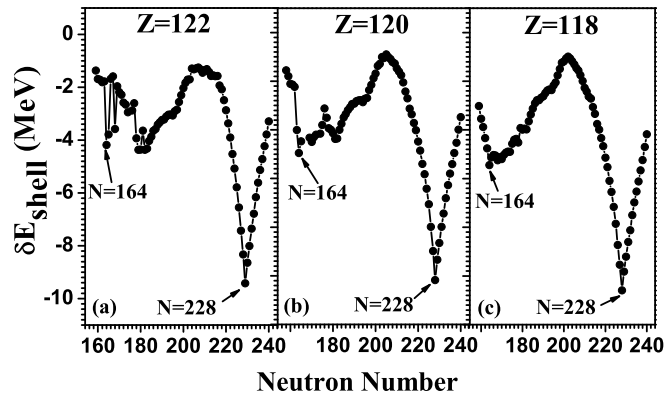}
\caption{Variation of shell correction $\delta$E$_{Shell}$ (in MeV) vs. N. The deep minima indicates magicity at N $=$ 164 and 228.}
\label{fig2}
\end{figure}

In search of new magic numbers in the superheavy region, we estimate shell correction values $\delta$E$_{Shell}$ for full isotopic chains of Z $=$ 122, 120 and 118 using NSM calculations shown in Fig. \ref{fig2}. The shell correction to energy $\delta$E$_{Shell}$ which is a minima at around shell closures, shows two minima with $\delta$E$_{Shell}$ values around 5 MeV and 10 MeV at N $=$ 164 and 228 as also found in our earlier work ~\cite{saxenaijmpe2018}. This shows very strong magic character which is in agreement with the other theoretical works \cite{Zhang2004,Lombard1976,Patra1999}. Since the magic nuclei are expected to have zero deformation with spherical shape, we find out the shape and deformations of the superheavy nuclei in this region to further validate the magic character of N=164 and 228. We trace energy minima corresponding to Nilsson deformation parameters namely axial deformation parameter $\beta$ and the shape parameter $\gamma$ while including the triaxial shapes also in addition to prolate and oblate shapes. \par

\begin{figure}[h]
\centering
\includegraphics[width=0.7\textwidth]{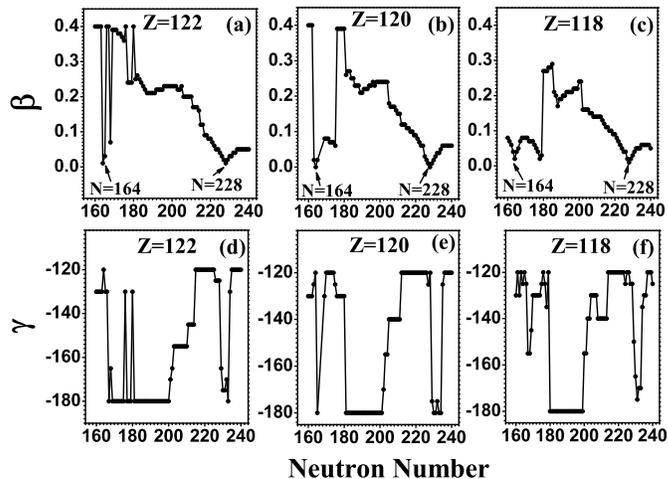}
\caption{Variation of $\beta_2$ and $\gamma$ vs. N for Z $=$ 122, 120 and 118 (using NSM) where the shapes are denoted by $\gamma$ $=$ -180$^o$ (oblate), -120$^o$ (prolate) and all other (triaxial).}
\label{fig3}
\end{figure}

Here we present for the first time a complete trace of equilibrium deformations and shapes of the isotopic chain of Z $=$ 122, 120 and 118 in Fig. \ref{fig3}. Nuclei in this region are found to be well deformed with $\beta$ ranging between 0.2 $-$ 0.4 as seen in Figs. \ref{fig3}(a), (b), (c) with a minima of zero deformation at N $=$ 228 confirming the magic character.  \par
The nuclear shapes are shown in the lower panels of Fig. \ref{fig3}, where we have plotted the shape parameter $\gamma$ vs N. We observe shape transitions from oblate ($\gamma$ $=$ -180$^o$) to triaxial (-120$^o$ $<$ $\gamma$ $<$ -180$^o$) while moving from A $=$ 299 $-$ 321 to A $=$ 322 $-$ 334. The $\beta$ values of RMF have not been included in this work as they show large deviations from NSM $\beta$ values which needs investigation although some the deviations are expected in RMF values as they do not include the triaxial shape in the calculations whereas NSM calculation derive the E minima considering all $\gamma$ values for all the shapes. For A $=$ 335 $-$ 348, prolate ($\gamma$ $=$ -120$^o$) minima is predominant. N $=$ 228 shows zero deformation which is obviously spherical in shape confirming the magic character. While undergoing the shape transitions, we observe that many nuclei appear to be potential candidates for shape coexistence which we would elaborate in our upcoming work. Shell correction to energy $\delta$E$_{shell}$ (see Fig. \ref{fig2}) varies from $\approx$ -10 MeV (at N $=$ 228) for spherical nucleus to upto 1 MeV (at mid shell) for well deformed nucleus points towards the shape transitions from  spherical (at N $=$ 228) to deformed (mid shell) which is evident in the plots of deformation and shapes and reaffirms the magic character of N $=$ 228.\par
\begin{figure}[h]
\centering
\includegraphics[width=0.7\textwidth]{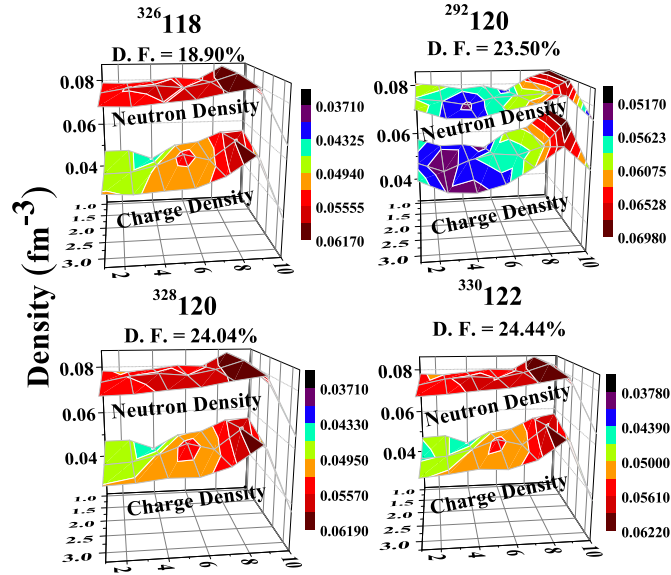}
\caption{Variation of charge density and neutron density for few selected nuclei $^{326}$118, $^{292}$120, $^{328}$120 and $^{330}$122.}
\label{fig4}
\end{figure}

Recently, central depletion in charge density named as 'bubble structure' has gain a lot of interest which has experimentally observed for the case of $^{34}$Si by Mutschler \textit{et al.} \cite{Mutschler2016}. In the light region, the central depletion is found due to depopulation of s state, whereas in superheavy region it is reported and investigated \cite{Berger2001,Decharge2003,Afanasjev2005,SinghSK2013,Li2016,Schuetrumpf2017,Saxenaplb2018,Saxenaplb2019} mainly due to large repulsive Coulomb field. With this in view, we investigate depletion in central density (bubble/semi-bubble structure) in the isotopes of Z $=$ 122, 120 and 118 using RMF+BCS approach. The depletion is found commonly in the superheavy nuclei. The prominent ones are displayed in Fig. \ref{fig4}, along with the neutron density. One can see that neutron density in all these selected nuclei depicts a constant variation in the interior of nucleus whereas charge density is found depleted near the center. It is remarkable to note here that central depletion is actually found in $^{292}$120 nucleus by RMF calculations inline with the other theoretical calculations \cite{Berger2001,Decharge2003,Afanasjev2005,SinghSK2013,Li2016,Schuetrumpf2017}. Other nuclei showing significant depletion are $^{326}$118, $^{328}$120 and $^{330}$122 from Z $=$ 118, 120 and 122 isotopic chains, respectively. In Fig. \ref{fig4}, we have expressed the depletion in charge density in terms of the depletion fraction (DF) which is based on value of density at the center and its maximum value. Therefore, central density depletion is  quantitatively represented by DF $=$ ($\rho_{max}-\rho_{c})/\rho_{max}$ \cite{Grasso2009}.

In order to check the effect of neutron number variation (if any) on depletion fraction, we have shown variation of D.F. with neutron number for isotopic chains of Z $=$ 122, 120 and 118 in the Table 1.

\begin{table}[htbp]
\caption{Depletion fraction (D.F.) calculated by RMF(TMA) for Z $=$ 122, 120 and 118 isotopes.}
\centering
\resizebox{0.4\textwidth}{!}{%
{\begin{tabular}{|c|c|c|c|}
 \hline
 \multicolumn{1}{|c|}{Neutron}&
 \multicolumn{3}{|c|}{Depletion fraction (D.F.)}\\
 \cline{2-4}
 \multicolumn{1}{|c|}{No.}&
 \multicolumn{1}{|c|}{Z$=$122}&
 \multicolumn{1}{|c|}{Z$=$120}&
 \multicolumn{1}{|c|}{Z$=$118}\\
\hline
205&24.52&24.04& 23.35\\
206&24.52&23.88& 23.39\\
207&24.52&23.66& 23.42\\
208&24.44&24.07& 23.50\\
209&24.64&23.99& 23.33\\
210&24.43&23.90& 23.08\\
211&24.47&23.73& 23.65\\
212&24.27&23.86& 22.86\\
213&23.97&23.90& 23.06\\
214&23.72&23.65& 22.64\\
215&23.76&23.51& 23.23\\
216&23.51&23.09& 22.43\\
217&23.42&23.51& 22.17\\
218&23.93&22.87& 22.20\\
219&23.04&22.74& 22.07\\
220&22.82&22.48& 21.85\\
221&22.52&22.20& 21.92\\
222&22.50&22.07& 21.66\\
223&22.84&22.45& 21.53\\
224&22.41&21.94& 21.29\\
225&21.94&21.67& 21.20\\
226&21.84&21.88& 21.27\\
227&21.37&21.27& 20.69\\
228&21.61&21.69& 21.07\\
229&21.78&21.59& 21.01\\
230&21.55&21.63& 21.18\\
\hline
\end{tabular}}}
\end{table}

\begin{figure}[h]
\centering
\includegraphics[width=0.7\textwidth]{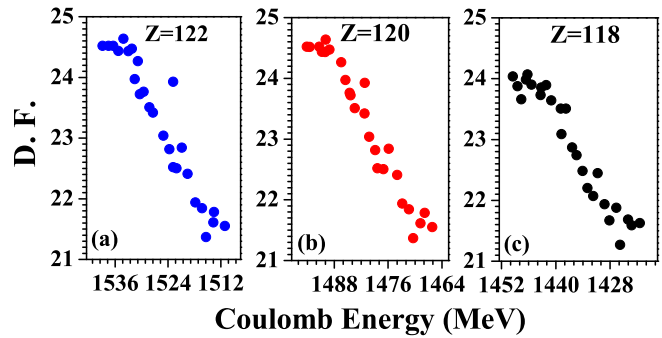}
\caption{Variation of depletion fraction (D.F.) with Coulomb energy for Z $=$ 122, 120 and 118 isotopes.}
\label{fig5}
\end{figure}

We find from the Table 1 that the depletion in the charge density decreases as the neutron number increases indicating effect of isospin (N$-$Z). In other words, as neutron number increases, the excess number of neutrons balance the Coulomb repulsion and consequently depletion fraction decreases. Such variation of D.F. with Coulomb energy is plotted in Fig. \ref{fig5} and evidently showing expected variation. This kind of variation in depletion fraction with respect to neutron number or Coulomb repulsion is expected to occur for all the superheavy nuclei which needs further investigations on the bubble phenomenon.

Towards the end of this subsection, we have given some ground state properties of Z $=$ 122, 120 and 118 isotopes in Table 2, which are found with larger binding energy per nucleon in their respective chains.

\begin{table}[!htbp]
\caption{Ground state properties viz. binding energy (B.E.), charge radii R$_c$, proton radii R$_p$, neutron radii R$_n$, matter radii R$_m$ and expected
shapes of Z $=$ 122, 120 and 118 isotopes are shown.}
\centering
\resizebox{0.6\textwidth}{!}{%
{\begin{tabular}{c|cc|cccc|c}
 \hline
 \hline\multicolumn{1}{c|}{Nuclei}&
 \multicolumn{2}{c|}{B. E.{(MeV)}}&
 \multicolumn{4}{c|}{RMF}&
 \multicolumn{1}{c}{Shape}\\
\cline{2-3}
\cline{4-7}
 \multicolumn{1}{c|}{}&
 \multicolumn{1}{c}{{RMF}}&
 \multicolumn{1}{c|}{{FRDM}}&
 \multicolumn{1}{c}{{R$_c$}}&
  \multicolumn{1}{c}{{R$_p$}}&
 \multicolumn{1}{c}{{R$_n$}}&
 \multicolumn{1}{c|}{R$_m$}&
 \multicolumn{1}{c}{{NSM}}\\
\hline
$^{298}${122}& 2088.09&           &6.56 &6.51 &  6.63 & 6.58 & Triaxial\\
$^{299}${122}& 2094.46&           &6.58 &6.54 &  6.66 & 6.61 & Oblate \\
$^{300}${122}& 2102.96&           &6.59 &6.54 &  6.68 & 6.62 & Oblate \\
$^{301}${122}& 2109.09&           &6.60 &6.55 &  6.69 & 6.63 & Oblate \\
$^{302}${122}& 2117.48&2113.69    &6.61 &6.56 &  6.70 & 6.64 &Triaxial \\
$^{303}${122}& 2123.40&2120.06    &6.64 &6.59 &  6.74 & 6.68 &Oblate \\
$^{304}${122}& 2131.48&2127.44    &6.65 &6.61 &  6.75 & 6.69 &Oblate \\
$^{305}${122}& 2137.39&2133.44    &6.66 &6.61 &  6.76 & 6.70 &Oblate \\
$^{306}${122}& 2145.06&2140.46    &6.67 &6.62 &  6.78 & 6.72 &Oblate \\
$^{307}${122}& 2150.76&2145.5     &6.68 &6.63 &  6.79 & 6.72 &Oblate \\
$^{308}${122}& 2158.39&2152.19    &6.69 &6.64 &  6.80 & 6.74 &Oblate \\
$^{309}${122}& 2163.59&2158.36    &6.73 &6.68 &  6.84 & 6.78 &Oblate \\
$^{310}${122}& 2171.02&2165.42    &6.73 &6.69 &  6.86 & 6.79 &Oblate \\
$^{311}${122}& 2176.04&2170.85    &6.74 &6.69 &  6.87 & 6.80 &Oblate \\
$^{312}${122}& 2183.21&2177.5     &6.78 &6.73 &  6.91 & 6.84 &Oblate \\
$^{313}${122}& 2187.87&2182.67    &6.76 &6.71 &  6.90 & 6.83 &Oblate \\
$^{314}${122}& 2195.07&2189.1     &6.79 &6.74 &  6.94 & 6.86 &Oblate \\
\hline
$^{290}${120}&2041.50&        &6.55&	6.50&	6.60&	6.56&Prolate \\
$^{291}${120}&2048.58&        &6.55&	6.50&	6.60&	6.56&Prolate \\
$^{292}${120}&2057.43&        &6.56&	6.51&	6.61&	6.57&Prolate \\
$^{293}${120}&2064.20&        &6.52&	6.47&	6.59&	6.54&Prolate \\
$^{294}${120}&2072.86&        &6.57&	6.52&	6.63&	6.59&Prolate \\
$^{295}${120}&2079.33&2077.82 &6.54&	6.49&	6.61&	6.56&Triaxial \\
$^{296}${120}&2087.83&2085.70 &6.58&	6.53&	6.65&	6.60&Triaxial \\
$^{297}${120}&2094.34&2092.17 &6.58&	6.53&	6.66&	6.61&Triaxial \\
$^{298}${120}&2102.29&2100.16 &6.59&	6.54&	6.68&	6.62&Triaxial \\
$^{299}${120}&2108.63&2106.33 &6.60&	6.55&	6.68&	6.63&Triaxial \\
$^{300}${120}&2116.35&2113.73 &6.60&	6.56&	6.70&	6.64&Triaxial \\
$^{301}${120}&2122.41&2119.70 &6.61&	6.56&	6.71&	6.65&Oblate \\
$^{302}${120}&2129.85&2126.77 &6.64&	6.59&	6.75&	6.69&Oblate \\
$^{303}${120}&2135.74&2132.47 &6.65&	6.60&	6.76&	6.69&Oblate \\
$^{304}${120}&2142.91&2139.18 &6.66&	6.61&	6.77&	6.71&Oblate \\
$^{305}${120}&2148.62&2143.86 &6.66&	6.61&	6.78&	6.72&Oblate \\
\hline
$^{284}${118}&2010.57& 2008.27 &6.31&	6.38&	6.25&	6.33& Triaxial\\
$^{285}${118}&2017.36& 2015.86 &6.29&	6.37&	6.24&	6.31& Triaxial\\
$^{286}${118}&2026.20& 2024.76 &6.30&	6.38&	6.25&	6.33& Triaxial\\
$^{287}${118}&2032.58& 2032.12 &6.31&	6.39&	6.26&	6.34& Triaxial\\
$^{288}${118}&2041.47& 2040.73 &6.50&	6.57&	6.45&	6.52& Triaxial\\
$^{289}${118}&2047.80& 2047.84 &6.50&	6.58&	6.45&	6.53& Triaxial\\
$^{290}${118}&2056.68& 2056.12 &6.51&	6.59&	6.46&	6.54& Triaxial\\
$^{291}${118}&2062.93& 2062.97 &6.51&	6.60&	6.46&	6.55& Triaxial\\
$^{292}${118}&2071.28& 2070.99 &6.52&	6.61&	6.47&	6.56&Triaxial \\
$^{293}${118}&2077.60& 2077.52 &6.53&	6.63&	6.48&	6.57&Triaxial \\
$^{294}${118}&2085.74& 2085.10 &6.54&	6.64&	6.49&	6.58&Prolate\\
$^{295}${118}&2091.32& 2091.77 &6.54&	6.65&	6.49&	6.59&Triaxial \\
$^{296}${118}&2099.27& 2099.12 &6.55&	6.67&	6.50&	6.60&Triaxial \\
$^{297}${118}&2104.69& 2105.02 &6.56&	6.68&	6.51&	6.61&Prolate\\
$^{298}${118}&2112.42& 2112.03 &6.57&	6.69&	6.52&	6.62&Oblate \\
$^{299}${118}&2117.67& 2117.69 &6.58&	6.70&	6.53&	6.63&Oblate \\
\hline
\end{tabular}}}
\end{table}

\clearpage
\subsection{Decay modes of nuclei with Z $=$ 122, 120 and 118}
Prediction of decay modes, decay products, Q-values and half lives are very much crucial to probe superheavy nuclei, and knowledge of these properties is essential for the detection of these superheavy nuclei in laboratory \cite{Hofmann2016,Oganessian2009,Bao2015,Wang2015,Niyti2015,Heenen2015,Santhosh2016,Budaca2016,Liu2017,Zhang2017}.
Here we present a systematic study of the decay properties of superheavy nuclei by calculating the $\alpha$-decay and spontaneous fission half-lives of Z $=$ 122, 120 and 118 isotopes within the range of A $=$ 284 $-$ 352 using RMF+BCS and NSM approaches. We have also calculated  Q-values and $log_{10}T_{\alpha}$ for the decay chains of superheavy nuclei $^{294}$118, $^{294}$117 and $^{293}$117, which have already been synthesized and their $\alpha$-decay chains have been reported \cite{Oganessian2010}, to validate our theories. To calculate log$_{10}T_{\alpha}$, we use recently reported modified Royer formula by Akrawy \textit{et al.} \cite{Akrawy2017}.
 \begin{equation}
 log_{10}T_{\alpha}(sec) = a + bA^{1/6}\sqrt{Z} + \frac{cZ}{\sqrt{Q_{\alpha}}}+ dI + eI^{2}
 \label{alpha}
\end{equation}
where I $=$ $\frac{N-Z}{A}$ and the constants a, b, c, d, and e are\\
\begin{table}[!htbp]
\centering
\resizebox{0.7\textwidth}{!}{%
{\begin{tabular}{|c|c|c|c|c|c|}
\hline
 \multicolumn{1}{|c|}{Nuclei (Z$-$N)}&
 \multicolumn{1}{|c|}{a}&
 \multicolumn{1}{|c|}{b}&
 \multicolumn{1}{|c|}{c}&
 \multicolumn{1}{|c|}{d}&
 \multicolumn{1}{|c|}{e}\\
 \hline
 $e-e$&-27.837&-0.9420&1.5343&-5.7004&8.785\\
 $o-e$&-26.801&-1.1078&1.5585&14.8525&-30.523\\
 $e-o$&-28.225&-0.8629&1.5377&-21.145&53.890\\
 $o-o$&-23.635&-0.891&1.404&-12.4255&36.9005\\
\hline
\end{tabular}}}
\end{table}

To calculate spontaneous fission half-lives (T$_{SF}$), We are using the formula of C. Xu $\textit{et al.}$ \cite{Xu2008} given below.
\begin{eqnarray}
T_{SF}(1/2) = exp[2\pi\{C_{0} + C_{1}A + C_{2}Z^2 + C_{3}Z^4 \nonumber\\ +  C_{4}(N-Z)^2 -(0.13323\frac{Z^2}{A^{1/3}} - 11.64)\}]
\label{SF}
\end{eqnarray}
The constants are $C_0$ $=$ -195.09227, $C_1$ $=$ 3.10156, $C_2$ $=$ -0.04386, $C_3$ $=$ 1.4030$\times10^{-6}$, and $C_4$ $=$ -0.03199.

In Table 3, we have shown results of Q$_{\alpha}$, $\log$T$_{\alpha}$ and possible decay mode computed by both theories (RMF and NSM) and compared with the experimental results \cite{Oganessian2010,Wang-Mass2017}. From Table 3, one can see that both the theories (RMF and NSM) are in reasonable agreement with the experimental data and hence provide certification for the prediction of decay properties of Z $=$ 122, 120 and 118 superheavy nuclei.\par

\begin{table*}[!htbp]
\caption{Comparison among $\alpha$-decay chains of $^{294}$118, $^{293}$117 and $^{294}$117 calculated by RMF and Mac-Mic theory with available experimental data \cite{Oganessian2010,Wang-Mass2017}.}
\centering
\resizebox{1.0\textwidth}{!}{%
{\begin{tabular}{c|ccc|c|cc|c|ccc}
 \hline
 \multicolumn{1}{c|}{Nuclei}&
 \multicolumn{3}{c|}{Q$_{\alpha}$(MeV)}&
 \multicolumn{1}{c|}{$\log$T(1/2)(sec.)}&
 \multicolumn{2}{c|}{$\log$T$_{\alpha}$(1/2)(sec.)}&
 \multicolumn{1}{c|}{$\log$T$_{SF}$(1/2)}&
 \multicolumn{3}{c}{Decay mode}\\
 \cline{2-4}
  \cline{6-7}
   \cline{9-11}
 \multicolumn{1}{c|}{}&
 \multicolumn{1}{c}{Expt.}&
 \multicolumn{1}{c}{RMF}&
 \multicolumn{1}{c|}{Mac-Mic}&
 \multicolumn{1}{c|}{Expt.}&
 \multicolumn{1}{c}{RMF}&
 \multicolumn{1}{c|}{Mac-Mic}&
 \multicolumn{1}{c|}{(sec.)}&
 \multicolumn{1}{c}{Expt.}&
 \multicolumn{1}{c}{RMF}&
 \multicolumn{1}{c}{Mac-Mic}\\
 \hline
 $^{294}${118}&11.84&11.15&13.63 &-3.16 &-0.80  & -5.96&8.48 &$\alpha$&$\alpha1$&$\alpha1$ \\
$^{290}${116}&11.00 &10.45&12.33&-2.10&0.32   & -4.03&3.80 &$\alpha$&$\alpha2$&$\alpha2$  \\
$^{286}${114}&10.37 &10.59&10.37&-0.85 &-0.71  & -0.12&0.38 &$\alpha$&$\alpha3$&$\alpha3$  \\
$^{282}${112}&10.17 &10.65&9.64&-3.04 &-1.51  &  1.19 &-1.89&SF  &SF       &SF \\
 \hline
 $^{293}${117}&11.03&10.91&13.23&-1.84 &-0.73&-5.79&8.18&$\alpha$ & $\alpha$ & $\alpha$\\
 $^{289}${115}&10.31&10.01&10.91&-0.65 & 1.05&-1.31&3.50&$\alpha$ & $\alpha$ & $\alpha$\\
 $^{285}${113}&10.01&10.30&10.07& 0.74 &-0.36& 0.26&0.07&$\alpha$ & $\alpha$ & $\alpha$\\
\hline
$^{294}${117}&10.81&10.75 &12.84 &-1.11  &0.62 &-3.64 &7.88 &$\alpha$ & $\alpha$ &  $\alpha$\\
$^{290}${115}&10.45&10.27 &10.99 &-1.80  &1.17 &-0.52 &3.21 &$\alpha$ & $\alpha$ &  $\alpha$\\
$^{286}${113}&9.79 &9.33  &9.72  &1.29   &3.02 &1.97  &-0.22&$\alpha$ &SF        & SF       \\
$^{282}${111}&9.64 &9.98  &9.28  &-0.29  &0.69 &2.52  &-2.49&$\alpha$ &SF        & SF       \\
$^{278}${109}&9.63 &9.93  &8.60  &0.88   &0.19 &3.83  &-3.68&$\alpha$ &SF        & SF       \\
$^{274}${107}&8.95 &8.34  &8.11  &1.73   &3.95 &4.69  &-3.86&SF       &SF        & SF       \\
\hline
\end{tabular}}}
\end{table*}

We have also looked into the possibility of $\beta$-decay which is investigated in superheavy nuclei by Fiset \textit{et al.} \cite{Fiset1972}, Karpov \textit{et al.} \cite{Karpov2012} and recently by Ikram \textit{et al.} \cite{Ikram2017}. We have used the empirical formula of Fiset and Nix \cite{Fiset1972} for estimating
the $\beta$-decay half-lives
\begin{equation}
T_{\beta}^{1/2} = (540 \times 10^5)\frac{m_{e}^{5}}{\rho_{d.s.}(W_{\beta}^{6} - m_{e}^{6})}
\label{beta}
\end{equation}

\begin{figure}[h]
\centering
\includegraphics[width=0.7\textwidth]{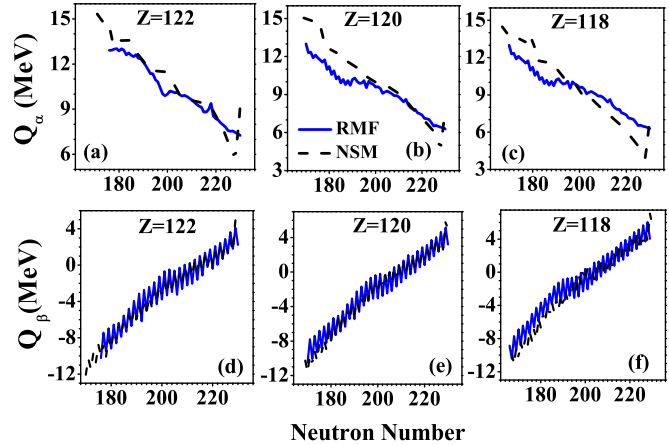}
	\caption{Variation of Q$_{\alpha}$ and Q$_{\beta}$ for considered isotopic chains as a function of neutron number N calculated by using RMF and NSM approaches.}
\label{fig6}
\end{figure}

Here, W$_{\beta}$ $=$ Q$_{\beta}$ + m$_{e}$, is the total maximum energy of the emitted $\beta$-particle in which
m$_{e}$ is rest mass and Q$_{\beta}$ $=$ BE(Z + 1,A) - BE(Z,A). $\rho_{d.s.}$ represents average density of states
in the daughter nuclei and is equal to $e^{-A/290} \times$ number of states within 1 MeV of ground states. For our purpose the values of number of states
are used as 2.73 and 8.6 for even and odd mass nuclei respectively taken from Ref.$~$ \cite{Seeger1965}. Fig. \ref{fig6} displays our calculated values (using RMF and NSM)  of Q$_{\alpha}$ and Q$_{\beta}$ for considered isotopic chains as a function of neutron number N, which show reasonable agreement. One can see that from the NSM calculations there are certain jumps in Q$_{\alpha}$ values at N $=$ 230, which are due to the magicity at N $=$ 228 as already reported in Ref. $~$\cite{saxenaijmpe2018} for the case of Z = 121 isotopes.

\begin{figure}[h]
\centering
\includegraphics[width=0.9\textwidth]{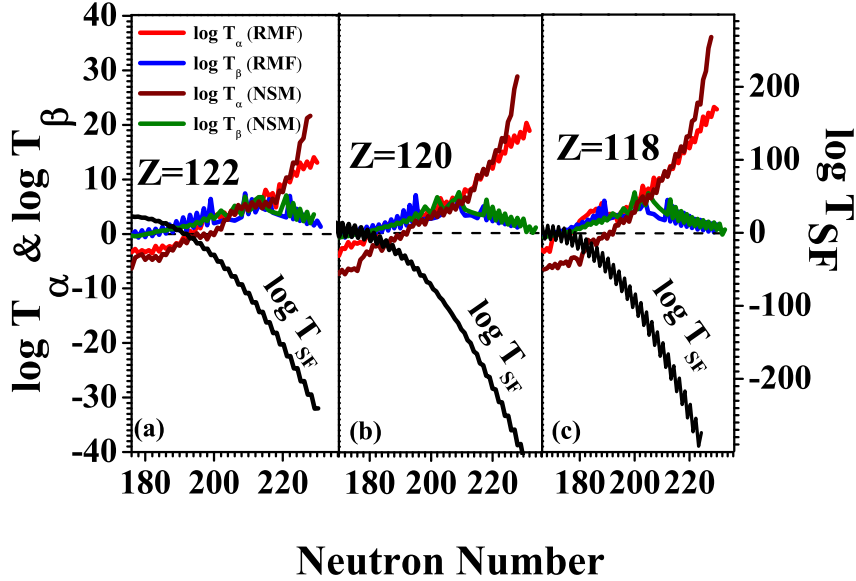}
	\caption{Variation of $log_{10}T_{\alpha}$, $log_{10}T_{\beta}^{1/2}$ (left scale) and log\,T$_{SF}$ (right scale) with respect to N for considered isotopic chains.}
\label{fig7}
\end{figure}

Fig. \ref{fig7} shows the varation of $log_{10}T_{\alpha}$ calculated using modified Royer formula \cite{Akrawy2017} (Eq. (\ref{alpha})) along with values of $log_{10}T_{\beta}^{1/2}$ calculated by formula of Fiset and Nix \cite{Fiset1972} (Eq. (\ref{beta})) for a comparison (left scale). We have also shown spontaneous fission half-life (T$_{SF}$) calculated by formula of Xu \textit{et al.} \cite{Xu2008} (Eq. (\ref{SF})) (right scale).

From Fig. \ref{fig7}, it is evident for the considered three isotopic chains (three panels of the figures) that $\alpha$-decay half-life ($log_{10}T_{\alpha}$) is lower than $\beta$-decay half-life $log_{10}T_{\beta}^{1/2}$ and spontaneous fission half-life (T$_{SF}$) for lower mass region which shows that the $\alpha$-decay is found to be favourable decay mode for A $\leq$ 314, 305 and 299 nuclei with Z $=$ 122, 120 and 118, respectively. Beyond these A values one can see a clear competition between $\alpha$-decay and $\beta$-decay for all these isotopic chains, however, for these A values (A $>$ 314, 305 and 299) spontaneous fission half-life (T$_{SF}$) becomes lower and decay through spontaneous fission becomes more favourable. Therefore, it may be concluded that chances of decay thorough $\beta$ are negligible in this superheavy domain. The nuclei with A $<$ 314, 305 and 299 with Z $=$ 122, 120 and 118 respectively, are found at the top of their respective plot of B.E./A (not shown here) and therefore are found most stable among full isotopic chains with B.E./A $\approx$ 7 MeV. One can see that this value of B.E./A is very close to the average of B.E./A (7.2 MeV) of around 80 nuclei which are experimentally known in this region so far.\par

Table 4, \& 5 show the values of Q$_{\alpha}$ and $\alpha$-decay half-life (T$_{\alpha}$) \cite{Akrawy2017} calculated by using RMF, NSM and FRDM \cite{Moller2012} approaches along with spontaneous fission half-life (T$_{SF}$) \cite{Xu2008}. The favorable decay modes predicted on the basis of comparison between T$_{\alpha}$ and T$_{SF}$ are also listed. These all data and modes are compared with available experimental data taken from Refs.$~$\cite{Oganessian2015,Wang-Mass2017} and are found in excellent match. It is notable from these tables that the nuclei between A $=$ 292 $-$ 302 (approximately 11 potential candidates) for Z $=$ 122 are found with long $\alpha$-decay chain for which our calculated $\alpha$-decay half-lives and predicted decay modes are in excellent agreement with available data from experiments \cite{Oganessian2015,Wang-Mass2017}. The decay chains of nuclei with 303 $\leq$ A $\leq$ 312 are terminated with SF after 4$\alpha$/3$\alpha$/2$\alpha$ . Therefore such nuclei are still far from the reach of experiments as of now. Nevertheless, nuclei $^{292-302}$122, $^{290-300}$120 and $^{288-298}$118 are found to
have enough potential to be observed or produced experimentally, because of their long $\alpha$-decay chains and high binding energy per nucleon.

\begin{table*}[!htbp]
\caption{Predictions on the modes of decay of $^{292-308}${122} superheavy nuclei and their decay products (decay-chain) by comparing the $\alpha$-decay half-lives
(sec) and the corresponding SF half-lives (sec). The half-lives are calculated using modified Royer's formula given by Akrawy $\textit{et al.}$ \cite{Akrawy2017} and
formula of Xu $\textit{et al.}$ for spontaneous fission \cite{Xu2008}.}
\centering
\resizebox{1.0\textwidth}{!}{%
{\begin{tabular}{c|cccc|c|ccc|c|cccc}
 \hline
 \hline\multicolumn{1}{c|}{Nuclei}&
 \multicolumn{4}{c|}{Q$_{\alpha}${(MeV)}}&
 \multicolumn{1}{c|}{logT{(1/2)}}&
 \multicolumn{3}{c|}{logT$_{\alpha}${(1/2)}}&
 \multicolumn{1}{c|}{logT$_{SF}${(1/2)}}&
 \multicolumn{4}{c}{Decay Mode}\\
\cline{2-9}
 \cline{11-14}
 \multicolumn{1}{c|}{}&
 \multicolumn{1}{c}{{Expt.}}&
 \multicolumn{1}{c}{{RMF}}&
 \multicolumn{1}{c}{{Mac-Mic}}&
 \multicolumn{1}{c|}{{FRDM}}&
 \multicolumn{1}{c|}{{Expt.}}&
  \multicolumn{1}{c}{{RMF}}&
 \multicolumn{1}{c}{{Mac-Mic}}&
 \multicolumn{1}{c|}{{FRDM}}&
 \multicolumn{1}{c|}{}&
 \multicolumn{1}{c}{{Expt.}}&
 \multicolumn{1}{c}{{RMF}}&
 \multicolumn{1}{c}{{Mac-Mic}}&
 \multicolumn{1}{c}{{FRDM}}\\
 \hline
$^{292}${122}&&13.52&15.99& -   & &-4.42&-8.53 &   -  & 20.72&    &$\alpha1$&$\alpha1$&-\\
$^{288}${120}&&12.36&15.51& -   & &-2.69&-8.30 &   -  & 13.42&    &$\alpha2$&$\alpha2$&- \\
$^{284}${118}&&13.46&14.97& -   & &-5.44&-7.99 &   -  & 7.54&    &$\alpha3$ &$\alpha1$ &- \\
$^{280}${116}&&12.96&12.64& -   & &-5.07&-4.44 &   -  & 3.01&    &$\alpha4$ &$\alpha2$ &- \\
$^{276}${114}&&12.45&10.34& -   & &-4.66&0.18  &  -  & -0.29&    &$\alpha5$ &SF&- \\
$^{272}${112}&&12.05&12.07& -   & &-4.44&-4.47 &   -  & -2.42&    &$\alpha6$&$\alpha2$&- \\
\hline
$^{293}${122}&&13.79&15.32 &- &   &-4.30  & -6.88&- & 22.10 &    &$\alpha1$&$\alpha1$ &-\\
$^{289}${120}&&12.07&15.04 &- &   &-1.45  & -6.98&- & 14.77 &    &$\alpha2$&$\alpha2$ &- \\
$^{285}${118}&&13.76&14.49 &- &   &-5.40  & -6.64&- & 8.87  &    &$\alpha3$&$\alpha3$ &- \\
$^{281}${116}&&12.72&13.53 &- &   &-4.05  & -5.56&- & 4.30  &    &$\alpha4$&$\alpha4$ &- \\
$^{277}${114}&&12.32&9.82  &- &   &-3.86  &  2.15&- &0.99   &     &$\alpha5$& SF      &- \\
$^{273}${112}&&11.83&11.38 &- &   &-3.46  & -2.48&- & -1.17 &    &$\alpha6$&$\alpha6$ &- \\
\hline
$^{294}${122}&&13.33&15.24 &14.52& &-4.11  & -7.43&-6.25 & 22.39&    &$\alpha1$&$\alpha1$&$\alpha1$  \\
$^{290}${120}&&12.06&14.94 &13.75& &-2.09  & -7.47&-5.45 & 15.06&    &$\alpha2$&$\alpha2$&$\alpha2$  \\
$^{286}${118}&&13.02&14.01 &13.05& &-4.66  & -6.45&-4.71 & 9.16&    &$\alpha3$& $\alpha3$&$\alpha3$  \\
$^{282}${116}&&12.83&14.53 &12.62& &-4.85  & -7.85&-4.44 & 4.59&    &$\alpha4$& $\alpha4$&$\alpha4$  \\
$^{278}${114}&&12.10&9.18  &11.76& &-4.00  & 3.47 &-3.26 & 1.27&    &$\alpha5$&  SF      &$\alpha5$  \\
$^{274}${112}&&11.66&10.75 &11.41& &-3.66  & -1.57&-3.11 & -0.88&    &$\alpha6$&$\alpha6$&$\alpha6$  \\
\hline
$^{295}${122}&&13.47& 14.91&14.49&& -3.75&-6.27 &-5.57 &23.09&    &$\alpha1$& $\alpha1$&$\alpha1$ \\
$^{291}${120}&&12.05& 14.87&13.87&& -1.44&-6.75 &-5.05 &15.74&    &$\alpha2$& $\alpha2$&$\alpha2$  \\
$^{287}${118}&&12.92& 13.98&12.96&& -3.87&-5.82 &-3.94 &9.81 &    &$\alpha3$& $\alpha3$&$\alpha3$  \\
$^{283}${116}&&12.80& 13.95&12.29&& -4.25&-6.33 &-3.21 &5.22 &    &$\alpha4$& $\alpha4$&$\alpha4$  \\
$^{279}${114}&&11.93& 9.89 &11.60&& -3.09& 1.91 &-2.36 &1.88 &    &$\alpha5$&  SF      &$\alpha5$  \\
$^{275}${112}&&11.43& 10.07&11.56&& -2.63& 0.70 &-2.92 &-0.30&    &$\alpha6$& $\alpha6$&$\alpha6$  \\
\hline
$^{296}${122}&&13.32&14.56&14.95  & &-4.13  & -6.37&-7.01   &23.34&    &$\alpha1$&$\alpha1$&$\alpha1$  \\
$^{292}${120}&&11.84&14.93&13.77  & &-1.65  & -7.50&-5.53   &15.99&    &$\alpha2$&$\alpha2$&$\alpha2$   \\
$^{288}${118}&&11.89&13.92&12.86  & &-2.36  & -6.34&-4.38   &10.06&    &$\alpha3$&$\alpha3$&$\alpha3$   \\
$^{284}${116}&&13.08&13.37&12.22  & &-5.38  & -5.92&-3.68   &5.47 &    &$\alpha4$&$\alpha4$&$\alpha4$   \\
$^{280}${114}&&11.89&10.71&11.05  & &-3.59  & -0.86&-1.69   &2.12 &    &$\alpha5$&$\alpha5$&$\alpha5$   \\
$^{276}${112}&11.91&11.22&9.58 &11.90  &-4.00 &-2.74  & 1.49 &-4.21   &-0.06&    &$\alpha6$& SF      &$\alpha6$   \\
\hline
$^{297}${122}&&13.39&14.59&14.93 & &-3.62  &-5.78&-6.34 &23.48&    &$\alpha1$& $\alpha1$ &$\alpha1$\\
$^{293}${120}&&11.91&14.68&13.65 & &-1.18  &-6.48&-4.69 &16.11&    &$\alpha2$& $\alpha2$ &$\alpha2$ \\
$^{289}${118}&&11.42&13.83&12.76 & &-0.70  &-5.59&-3.59 &10.17&    &$\alpha3$& $\alpha3$ &$\alpha3$ \\
$^{285}${116}&&13.43&13.20&11.85 & &-5.45  &-5.03&-2.31 &5.55 &    &$\alpha4$& $\alpha4$ &$\alpha4$ \\
$^{281}${114}&&11.59&10.47&10.58 & &-2.39  &0.31 &0.03  &2.19 &    &$\alpha5$& $\alpha5$ &$\alpha5$ \\
$^{277}${112}&11.62&11.03&9.84 &12.13 &-3.07 &-1.76  &1.30 &-4.15 &-0.01&    &$\alpha6$&  SF       &$\alpha6$ \\
 \hline
$^{298}${122}&&13.06&14.50&15.16& &-3.68  &-6.31&-7.39 &23.59&    &$\alpha1$&$\alpha1$ &$\alpha1$\\
$^{294}${120}&&11.70&14.45&13.49& &-1.38  &-6.76&-5.06 &16.20&    &$\alpha2$&$\alpha2$ &$\alpha2$ \\
$^{290}${118}&&11.38&13.93&12.68& &-1.25  &-6.40&-4.07 &10.25&    &$\alpha3$&$\alpha3$ &$\alpha3$ \\
$^{286}${116}&&12.27&12.97&11.68& &-3.84  &-5.21&-2.56 &5.63&     &$\alpha4$&$\alpha4$ &$\alpha4$ \\
$^{282}${114}&&11.92&10.27&9.96& &-3.70   &0.23 & 1.07 &2.25&     &$\alpha5$&$\alpha5$ &$\alpha5$ \\
$^{278}${112}&11.31&10.87&10.10&12.16&-2.70 &-1.97  &0.00 &-4.79 &0.04&     &$\alpha6$&$\alpha6$ &$\alpha6$ \\
\hline
$^{299}${122}&&13.16&13.74&14.63 & &-3.22  &-4.32&-5.88 &23.41&    &$\alpha1$& $\alpha1$&$\alpha1$\\
$^{295}${120}&&11.99&14.63&13.45 & &-1.39  &-6.43&-4.36 &16.02&    &$\alpha2$& $\alpha2$&$\alpha2$ \\
$^{291}${118}&&11.08&13.67&12.55 & &0.08   &-5.34&-3.20 &10.06&    &$\alpha3$& $\alpha3$&$\alpha3$ \\
$^{287}${116}&&11.69&12.77&11.21 & &-2.00  &-4.24&-0.88 &5.43 &    &$\alpha4$& $\alpha4$&$\alpha4$ \\
$^{283}${114}&&12.04&10.31&9.82 & &-3.38   & 0.70&2.05  &2.05 &    &$\alpha5$& $\alpha5$&$\alpha5$/SF \\
$^{279}${112}&11.04&10.80&9.95 &11.66 &-2.30 &-1.25  & 0.97&-3.20 &-0.17&    &$\alpha6$&SF        &$\alpha6$ \\
\hline
$^{300}${122}&&13.16&13.21&14.72& &-3.92  & -4.01&-6.72 &23.11&    &$\alpha1$&$\alpha1$&$\alpha1$ \\
$^{296}${120}&&11.47&14.51&13.59& &-0.89  & -6.90&-5.29 &15.70&    &$\alpha2$&$\alpha2$&$\alpha2$  \\
$^{292}${118}&&11.28&13.44&12.39& &-1.05  & -5.57&-3.52 &9.72 &    &$\alpha3$&$\alpha3$&$\alpha3$  \\
$^{288}${116}&&10.87&12.76&11.20& &-0.71  & -4.85&-1.49 &5.07 &    &$\alpha4$&$\alpha4$&$\alpha4$  \\
$^{284}${114}&10.80&11.80&10.27&9.52&-2.48  &-3.48  & 0.18 &2.29  &1.67 &    &$\alpha5$&$\alpha5$&SF  \\
$^{280}${112}&10.73&10.80&9.86 &11.08&-2.30 &-1.82  & 0.61 &-2.49 &-0.57&    &$\alpha6$&SF       &$\alpha5$ \\
\hline
\end{tabular}}}
\end{table*}
\begin{table*}[!htbp]
\caption{Table 4 Cont...}
\centering
\resizebox{1.0\textwidth}{!}{%
{\begin{tabular}{c|cccc|c|ccc|c|cccc}
 \hline
 \hline\multicolumn{1}{c|}{Nuclei}&
 \multicolumn{4}{c|}{Q$_{\alpha}${(MeV)}}&
 \multicolumn{1}{c|}{logT{(1/2)}}&
 \multicolumn{3}{c|}{logT$_{\alpha}${(1/2)}}&
 \multicolumn{1}{c|}{logT$_{SF}${(1/2)}}&
 \multicolumn{4}{c}{Decay Mode}\\
\cline{2-9}
 \cline{11-14}
 \multicolumn{1}{c|}{}&
 \multicolumn{1}{c}{{Expt.}}&
 \multicolumn{1}{c}{{RMF}}&
 \multicolumn{1}{c}{{Mac-Mic}}&
 \multicolumn{1}{c|}{{FRDM}}&
 \multicolumn{1}{c|}{{Expt.}}&
  \multicolumn{1}{c}{{RMF}}&
 \multicolumn{1}{c}{{Mac-Mic}}&
 \multicolumn{1}{c|}{{FRDM}}&
 \multicolumn{1}{c|}{}&
 \multicolumn{1}{c}{{Expt.}}&
 \multicolumn{1}{c}{{RMF}}&
 \multicolumn{1}{c}{{Mac-Mic}}&
 \multicolumn{1}{c}{{FRDM}}\\
 \hline
$^{301}${122}&&13.55&13.42&14.47& &-4.00  & -3.75&-5.64 &22.84&    &$\alpha1$&$\alpha1$&$\alpha1$ \\
$^{297}${120}&&11.39&13.73&13.65& &-0.03  & -4.91&-4.76 &15.43&    &$\alpha2$&$\alpha2$&$\alpha2$  \\
$^{293}${118}&11.92&11.26&13.58&12.34&-3.00 &-0.39  & -5.21&-2.79 &9.44 &    &$\alpha3$&$\alpha3$&$\alpha3$  \\
$^{289}${116}&11.10&10.34&12.55&11.10&-2.70 &1.27   & -3.84&-0.65 &4.79 &    &$\alpha4$&$\alpha4$&$\alpha4$  \\
$^{285}${114}&10.56&11.39&10.27&9.44 &-0.89 &-1.99  & 0.78&3.14  &1.39 &$\alpha$ &$\alpha5$&$\alpha5$&SF  \\
$^{281}${112}&10.46&10.90&9.75&10.24 &-1.00&-1.51  & 1.50&0.16  &-0.85&$\alpha$    &$\alpha6$& SF& SF \\
\hline
$^{302}${122}&&13.10&13.57 &14.77& &-3.84  & -4.74&-6.85&21.93&    &$\alpha1$&$\alpha1$& $\alpha1$\\
$^{298}${120}&&11.22&13.07 &13.24& &-0.31  & -4.35&-4.68&14.49&    &$\alpha2$&$\alpha2$& $\alpha2$ \\
$^{294}${118}&11.84&11.15&13.63&12.37 &-3.16 &-0.80  & -5.96&-3.52&8.48 &$\alpha$&$\alpha3$&$\alpha3$ &$\alpha3$\\
$^{290}${116}&11.00&10.45&12.33&11.06 &-2.10&0.32   & -4.03&-1.20&3.80 &$\alpha$&$\alpha4$&$\alpha4$&$\alpha4$  \\
$^{286}${114}&10.37&10.59&10.37& 9.48&-0.85 &-0.71  & -0.12&2.37&0.38 &$\alpha$&$\alpha5$&$\alpha5$&SF  \\
$^{282}${112}&10.17&10.65&9.64&9.43 &-3.04 &-1.51  &  1.19&1.80 &-1.89&SF  &SF       &SF &SF\\
\hline
$^{303}${122}&&13.53&13.56&14.57 & &-3.99  & -4.05&-5.84 &21.63&    &$\alpha1$&$\alpha1$&$\alpha1$ \\
$^{299}${120}&&10.94&12.91&13.74 & &1.05   & -3.38&-4.95 &14.19&    &$\alpha2$&$\alpha2$&$\alpha2$  \\
$^{295}${118}&11.70&11.13&13.22&11.92 &-2.00 &-0.10  & -4.57&-1.92 &8.18 &    &$\alpha3$&$\alpha3$&$\alpha3$  \\
$^{291}${116}&10.89&10.37&12.48&11.02&-1.72 &1.16   & -3.72&-0.48 &3.51 & $\alpha$&$\alpha4$&$\alpha4$&$\alpha4$  \\
$^{287}${114}&10.16&9.83 &10.03&9.40 &-0.23 &1.96   & 1.41 &3.23 &0.08 &$\alpha$&SF       &SF &SF        \\
$^{283}${112}&9.94 &10.42&9.80 &9.13 &0.61 &-0.35  & 1.34 &3.32 &-2.19& $\alpha$   &SF       &SF &SF        \\
\hline
$^{304}${122}&&13.17&13.55&14.59 &&-4.01 & -4.74&-6.59&20.02&    &$\alpha1$&$\alpha1$&$\alpha1$ \\
$^{300}${120}&&10.77&12.88&13.69 &&0.79  & -4.01&-5.56&12.56&    &$\alpha2$&$\alpha2$&$\alpha2$  \\
$^{296}${118}&&10.91&12.78&12.28 &&-0.23 & -4.40&-3.37&6.53 &    &$\alpha3$&$\alpha3$&$\alpha3$  \\
$^{292}${116}&10.77&10.45&12.57&10.82 &-1.62&0.30  & -4.56&-0.65&1.82 &$\alpha$&$\alpha4$&$\alpha4$&$\alpha4$  \\
$^{288}${114}&10.07&9.48 &9.73 &9.17  &-0.12&2.32  & 1.60 &3.28&-1.63&$\alpha$    &SF&     SF &     SF          \\
$^{284}${112}&9.60&10.16&9.99 &8.97 &-0.98&-0.29 & 0.17&3.18 &-3.93&  SF  &SF&     SF&SF  \\
\hline
$^{305}${122}&&13.32&13.56&14.56&&-3.62&-4.07&-5.85&19.72&    &$\alpha1$&$\alpha1$&$\alpha1$ \\
$^{301}${120}&&10.57&12.33&13.62&&1.99 &-2.21&-4.76&12.26&    &$\alpha2$&$\alpha2$&$\alpha2$  \\
$^{297}${118}&&10.59&13.27&12.38&&1.27 &-4.69&-2.93&6.22 &    &$\alpha3$&$\alpha3$&$\alpha3$  \\
$^{293}${116}&10.68&10.51&12.25&10.78&-1.10&0.78&-3.27&0.10&1.52 &$\alpha$&$\alpha4$&$\alpha4$&$\alpha4$  \\
$^{289}${114}&9.97&9.61 &9.69 &9.06&0.38&2.59 &2.35  &4.28&-1.93&$\alpha$&SF&     SF &SF \\
$^{285}${112}&9.32&9.11 &9.66 &8.79&1.51&3.34 &1.72  &4.40&-4.23&$\alpha$&SF&     SF &SF \\
\hline
$^{306}${122}&&13.09&13.62&14.61&&-3.90 &-4.91&-6.66&17.41&    &$\alpha1$&$\alpha1$&$\alpha1$ \\
$^{302}${120}&&10.86&12.10&13.56&&0.49  &-2.43&-5.36&9.92 &    &$\alpha2$&$\alpha2$&$\alpha2$  \\
$^{298}${118}&&9.97 &13.48&12.49&&2.25  &-5.77&-3.85&3.86 &    &$\alpha3$&$\alpha3$&$\alpha3$  \\
$^{294}${116}&&10.39&11.91&10.90&&0.41  &-3.23&-0.89&-0.87&    &SF       &$\alpha3$&$\alpha3$   \\
$^{290}${114}&&9.52 &9.56 &8.84 &&2.16  &2.05 &4.31 &-4.36&    &SF       &SF       &$\alpha3$/SF\\
$^{286}${112}&&8.98 &9.38 &8.62 &&3.10  &1.87 &4.29 &-6.68&    &SF       &SF       &$\alpha3$/SF\\
\hline
$^{307}${122}&&13.28&13.35&15.27&&-3.56& -3.70&-7.03&17.11&    &$\alpha1$&$\alpha1$&$\alpha1$ \\
$^{303}${120}&&10.26&12.29&13.52&&2.82 & -2.14&-4.60&9.62 &    &$\alpha2$&$\alpha2$&$\alpha2$  \\
$^{299}${118}&&10.03&12.75&12.51&&2.77 & -3.70&-3.22&3.56 &    &$\alpha3$&$\alpha3$&$\alpha3$  \\
$^{295}${116}&&10.27&12.59&10.87&&1.40 & -3.98&-0.15&-1.17&    &SF&    $\alpha3$   & SF \\
$^{291}${114}&&9.67 &8.91 &8.76&&2.38 & 4.75 & 5.25&-4.66&     &SF&SF              & SF    \\
$^{287}${112}&&8.86 &9.47 &8.54&&4.13 & 2.25 & 5.23&-6.98&     &SF&SF              & SF   \\
\hline
$^{308}${122}&&12.82&13.05&15.29&&-3.40&-3.86&-7.80 &14.08&    &$\alpha1$&$\alpha1$&$\alpha1$ \\
$^{304}${120}&&10.09&12.45&13.55&&2.56& -3.22&-5.38 &6.57 &    &$\alpha2$&$\alpha2$&$\alpha2$  \\
$^{300}${118}&&10.00&12.11&12.51&&2.11& -3.09&-3.93 &0.48 &    &SF&$\alpha1$       &$\alpha2$  \\
$^{296}${116}&&9.99 &13.30&11.18&&1.46& -6.04&-1.61 &-4.28&    &SF&$\alpha2$       & SF  \\
$^{292}${114}&&9.54 &8.24 &8.32&&2.06& 6.38 & 6.08&-7.79&      &SF& SF             & SF \\
$^{288}${112}&&8.72 &9.51 &8.59&&3.90& 1.45 & 4.36&-10.14&     &SF&SF             &  SF \\
\hline
\hline
\end{tabular}}}
\end{table*}

\clearpage
\section{Summary}
We have employed two completely different approaches relativistic mean-field (RMF) plus BCS approach and the Macroscopic-Microscopic approach with Nilsson Strutinsky
prescription for an extensive and systematic study of even and odd isotopes of Z $=$ 122, 120 and 118 (A $=$ 284 $-$ 352). We investigate ground state properties
such as binding energy, separation energy, shell correction, deformation, shape, radii and charge density, and compared our results with the available results of FRDM which show good agreement with our calculations. A complete trace of separation energy, shell correction, deformation and shape is presented which shows the strong evidences for magicity in N $=$ 228. We predict the central depletion in charge density for few isotopes of Z $=$ 122, 120 and 118 which is due to strong
Coulomb repulsion due to a large number of protons. Depletion fraction is computed which is found to be stronger towards the neutron deficient side and decreases with
decreasing isospin and consequently Coulomb energy. We have investigated the decay properties of a series of all these isotopes and presented our results on the possible modes of decay by comparing the $\alpha$-decay half-lives, $\beta$-decay half-lives and the corresponding SF half-lives. We found almost no possibility of $\beta$-decay
in the full isotopic chains of these nuclei and the nuclei with A $>$ 312 are found favorable to decay through spontaneous fission whereas those with A $<$ 312
are expected to decay through $\alpha$. Nuclei with A $=$ 292 $-$ 302 are found with long $\alpha$-decay chains and hence the potential candidates for future experiments.
With this study, we expect to have moved a forward step into the unknown territory of superheavy nuclei for the future experimental inputs.

\section*{Acknowledgements}
Authors G.S. and M.A. acknowledge the support by SERB for YSS/2015/000952 and WOS-A schemes respectively.

\end{document}